\documentclass{article}
\usepackage{arxiv}

\usepackage[utf8]{inputenc}
\usepackage[T1]{fontenc}
\usepackage{hyperref}
\usepackage{url}
\usepackage{booktabs}
\usepackage{amsfonts}
\usepackage{nicefrac}
\usepackage{microtype}
\usepackage{lipsum}
\usepackage{graphicx}
\usepackage{algorithm}
\usepackage{algpseudocode}
\usepackage{color}
\usepackage{natbib} 
\usepackage{doi}
\usepackage{amsmath}
\usepackage{cleveref}
\usepackage{amssymb}
\usepackage{bbm}
\usepackage{mathtools}

\usepackage{amsthm}

\newtheorem{theorem}{Theorem}[section]

\theoremstyle{definition}

\newtheorem{definition}[theorem]{Definition}

\title{Identifying Chemicals in Metabolomic Samples Through Dimensionality Reduction}

\author{
\href{https://orcid.org/0000-0003-2893-9469}{\includegraphics[scale=0.06]{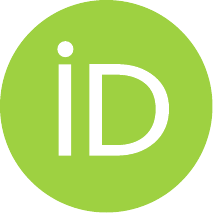}\hspace{1mm}Emile Anand}\thanks{This work was done while the author was visiting the Niels Bohr Institute at the University of Copenhagen in Denmark, and the research was funded by a Caltech Summer Undergraduate Research Fellowship (SURF) in 2020.} \\
	Computing and Mathematical Science\\
	California Institute of Technology\\
	Pasadena, CA, 91125 \\
	\texttt{eanand@caltech.edu}
    \and
\href{https://orcid.org/0000-0003-3780-6801}{\includegraphics[scale=0.06]{orcid.pdf}\hspace{1mm}\textbf{Charles Steinhardt}} \\
	Cosmic Dawn Center, Niels Bohr Institute\\
	University of Copenhagen\\
	Copenhagen, Denmark \\
	\texttt{steinhardt@nbi.ku.dk} \and
    \href{https://orcid.org/0000-0002-4663-8742}{\includegraphics[scale=0.06]{orcid.pdf}\hspace{1mm}\textbf{Martin Hansen}} \\
    Department of Environmental and Resource Engineering\\
	The Technical University of Denmark\\
	Lyngby, Denmark \\
	\texttt{marthan@dtu.dk}
}

\date{}

\renewcommand{\shorttitle}{Identifying Chemicals Through Dimensionality Reduction}

\hypersetup{
pdftitle={Identifying Chemicals Through Dimensionality Reduction},
pdfsubject={TCS},
pdfauthor={Emile T. Anand},
pdfkeywords={Expander Graphs · Random Walks · Sticky Random Walk · Pseudorandomness · Derandomization}
}

\begin{document}
\maketitle

\begin{abstract}
    The process of determining contaminants in sources of water has evolved with the complexity of the contaminants due to pesticides and heavy metals. The routine procedure to determine whether a sample of water is safe to consume is \emph{targeted analysis} which searches for specific substances from some known list; however, as any such list of contaminants grows in structural complexity, any such list is, unfortunately, non-exhaustive: we do not explicitly know which substances should be on this list. This raises the following fundamental issue: before experimentally determining which substances are contaminants, how do we answer the sampling problem of identifying all the substances in the water? Here, we present an approach that builds on the work of \citet{kruvequantification}, which uses non-targeted analysis by developing a random-forest regression model to predict the chemical formulas of each molecule in a sample, as well as their respective concentrations. This work utilizes techniques from dimensionality reduction and linear decompositions to present a more accurate unsupervised model using data from the European Massbank Metabolome Library \citep{oberacher:hal-01410982} to identify chemical formulas in new samples.

\end{abstract}


\newpage

\section{Introduction}
    There is an undeniable need to provide safe water across the world. Scientists currently use targeted analysis to search for water contaminants. A targeted analysis is a narrow search performed on a sample to identify specific substances. However, this introduces a statistical sampling problem: while scientists know how to test for and filter specific substances, how do they \emph{a priori} determine which substances these tests should encompass? An emerging method to analyze samples of water is untargeted analysis \citep{kruvequantification,alder2006residue,cajka2016toward}, which is when a broad search is performed on the water sample to detect both known and unknown chemicals using data from mass spectrometry. Untargeted analysis in this field necessitates a data-driven approach that utilizes machine learning to learn the function that maps data from a mass spectrometer to a chemical formula. Therefore, the onus of this problem of ensuring safe water access and identifying the composition of a metabolome (the set of chemicals in a biological sample) is to utilize a data-driven approach to learn this function and use it to identify and predict the concentration of each chemical in the metabolomic landscape.
    
    In 2020, \citet{kruvequantification} attempted to solve this problem by developing a hyperparamaterized regularized random forest-regressor. A random forest regression model is a discrete machine learning model that trains an ensemble of randomized regression trees on a random subset of the data. The model estimates values by averaging predictions over each random tree regressor. Liigand et. al.'s model achieved a strong regularization condition (a means to prevent over-fitting the model on the data) by enforcing a maximum leaf size of 120 nodes on each tree \citep{kruvequantification}. Therefore, upon input of data from a mass-spectrometer, the random forest regression model \citep{ChenGuestrin2016} discretized the output space to estimate the decision boundaries; in doing so, it learned some fragments of the underlying relationships between each chemical and its mass spectrum \citep{guptacryogenic}. Liigand et. al.'s hyperparameterized model produces a mean error of 8.8\% \citep{kruvequantification}, which can potentially be improved since discrete models ignore possible continuous relations between the mass spectra and the profile of the corresponding metabolite. This is provable using a t-stochastic neighbor embedding (t-SNE) argument:
    
    t-SNE \citep{vanDerMaaten2008} is a popular tool for visualizing high-dimensional data. It creates a probability distribution for each point (using a normalized Gaussian centered at that point) and minimizes the Kullback-Leibler divergence between this distribution and the distribution of a set of points in $\mathbb{R}^2$ using stochastic gradient descent algorithm. In $\mathbb{R}^2$, T-SNE tends to form clusters of points such that each cluster can be qualitatively described using a single variable; therefore, the number of clusters provide clues to the approximate minimum number of variables needed to describe the data \citep{vanDerMaaten2008}. Figure 1 displays a t-SNE plot that is color-coded with the presence of Chlorine atoms (here, chlorine-containing compounds are black and non-chlorine containing compounds are yellow). Figure 1b shows that there exists a reliable 3-variable clustering of substances in reduced-parameter form due to the presence of 3 clusters. Observe that the variance in the colored labels is reasonable since the clusterings cannot distinguish between similar and dissimilar data points (the t-SNE algorithm simply does not check for this). Thus, our approach to solving this problem can potentially produce a mean ground-truth error of ~0 since we do not make assumptions about the distribution \citep{waggoner2021modern}.

    \begin{center}
\includegraphics[scale=0.28]{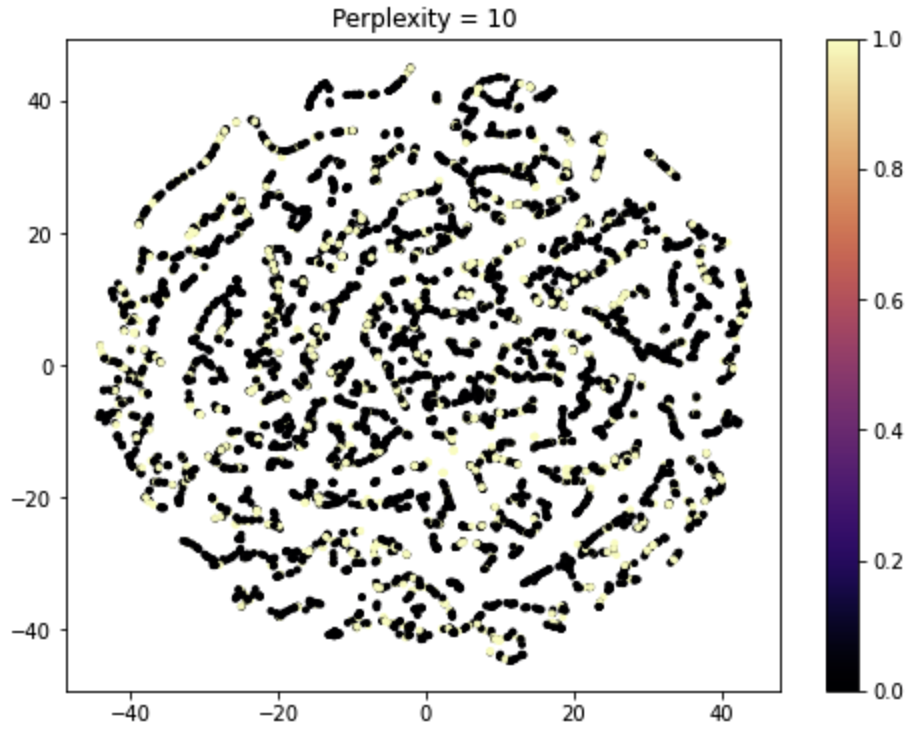}
\includegraphics[scale=0.28]{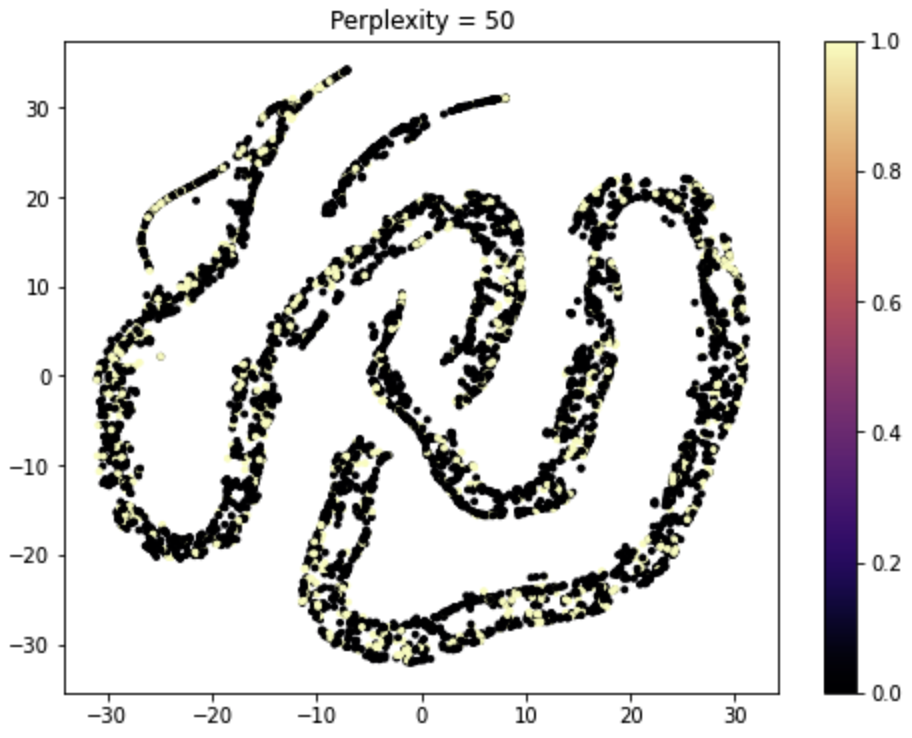}
\includegraphics[scale=0.28]{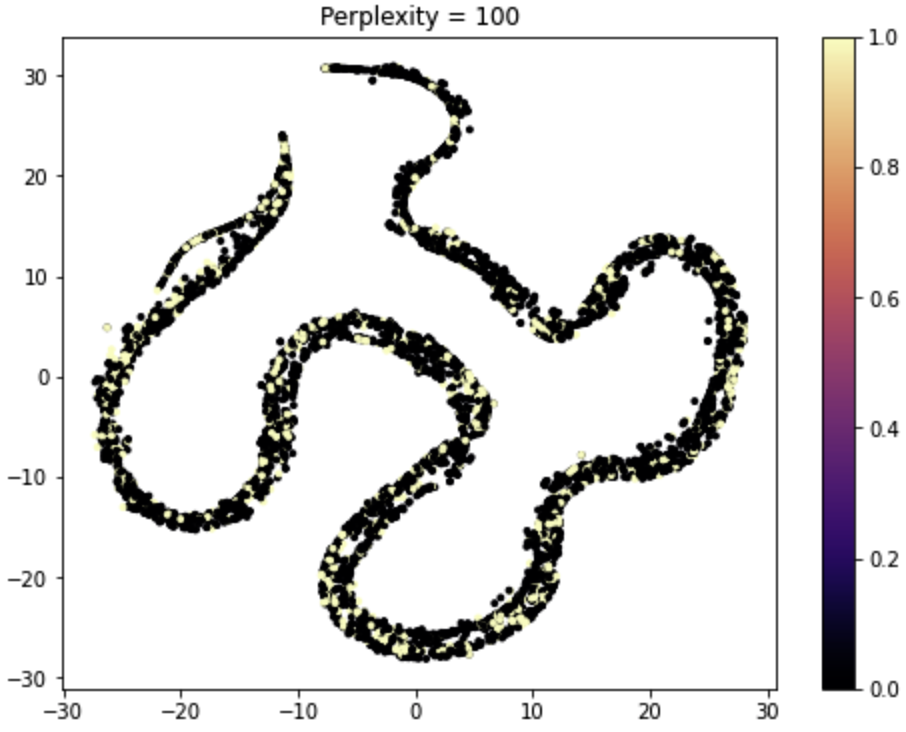}
\textbf{\newline Figure 1:} t-SNE applied to objects with perplexities of 10, 50, and 100, trained on 500 iterations. Low perplexity values form localized and complex structures. High perplexity values form global  structures with more significant clusterings.\end{center}

\section{Preliminaries}

We provide the following essential definitions.

\begin{definition}[KL divergence]
    For discrete probability distributions $P$ and $Q$ defined on the same sample space $\mathcal{X}$, the Kullback-Lieblerg (KL) divergence is given by
    \begin{equation}
        D_{\mathrm{KL}}(P\|Q) = -\sum_{x\in\mathcal{X}}P(x) \log\frac{Q(x)}{P(x)}
    \end{equation}
\end{definition}

The t-SNE model is defined through the KL-divergence score \citep{vanDerMaaten2008}.

\paragraph{\textbf{t}-\textbf{S}tochastic \textbf{N}eighbor \textbf{E}mbedding (t-SNE).} Given a set of $N$ high-dimensional objects $x_1,\dots,x_N$, t-SNE first computes probability $p_{ij}$ where $p_{ij} = \frac{p_{i|j}+p_{j|i}}{2N}$. Here,
\begin{equation}
    p_{i|j} = \begin{cases}\frac{\exp(-\|x_i - x_j\|^2/2\sigma_i^2)}{\sum_{k\neq i}\exp(-\|x_i-x_k\|^2/2\sigma_i^2)}, & i\neq j \\ 0, & \text{otherwise}\end{cases}
\end{equation}
Next, t-SNE learns a $d$-dimensional map $y_1,\dots,y_N$ (with $y_i \in \mathbb{R}^d$) where the similarity $q_{ij}$ between points $y_i$ and $y_j$ is given by $\mathrm{KL}(P\|Q)=\sum_{i\neq j}p_{ij}\log\frac{p_{ij}}{q_{ij}}$. Here, 
\begin{align}
    q_{ij} = \begin{cases}
        \frac{(1+\|y_i-y_j\|^2)^{-1}}{\sum_k \sum_{l\neq k} (1+\|y_k - y_l\|^2)^{-1}}, & i\neq j \\
        0, & \text{otherwise}
    \end{cases},
\end{align}
where the minimization of the KL divergence is performed using gradient descent \citep{lin2023online,lin2023online2,pmlr-v247-lin24a}.

\section{Methods} \label{sec:Methods}
This section describes the data-processing techniques and encoding chemical signals into SMILE strings.
    \subsection{Data Acquisition and Preprocessing}
      The ideal machine-learning model that reverse-engineers the process of mass spectrometry takes in a mass spectrum as an input and predicts the name of the substance corresponding to the spectra. Before constructing such a model, it is important that we only train it on qualitatively clean data. This necessitates the need to refine it.
      
      The Human Metabolome Database and the European MassBank Database use a mass-spectrometer to generate chromatograms and mass spectra of 1755357 peaks corresponding to 85582 various substances. Each spectra contains numerous peaks, and each peak represents a fragment of the ionized sample deflected in the mass spectrometer (as shown in Figure 2). The height of the peak is the relative intensity of the fragment in the sample, and the location of the peak is the mass-to-charge ratio of the fragment. To ensure that the data-points we would choose were of high-quality, we only extracted chemicals with more than 3 peaks in their corresponding spectrum in which each peak exceeded a predetermined threshold signal-to-noise ratio (SNR). In order to compare different spectra, we discretized the mass-charge values by binning the data. While binning leads to a small loss of information, it was essential for this problem since it is otherwise impossible to compare continuous spectra and it standardizes the different sub-datasets in the overall database, thus accounting for differences in how the data was posted.\\
      
    \includegraphics[scale=0.55]{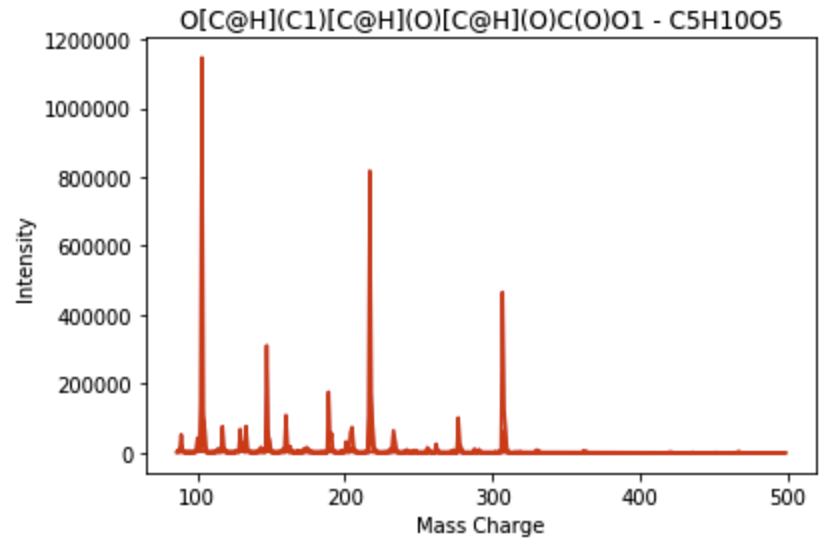}
    \includegraphics[scale=0.55]{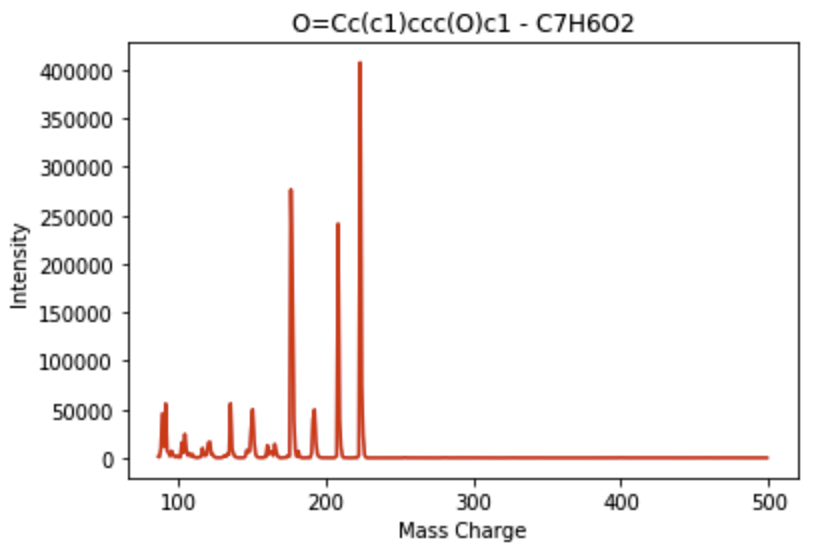}
    \begin{center}
        \textbf{Figure 2:} Visualizations of the spectrum of the compounds $C_{5}H_{10}O_{5}$ and $C_{7}H_{6}O_{2}$. The data was acquired from the European MassBank and the Human Metabolome Database, and the figures were generated through Python's Matplotlib graphics engine.
    \end{center}

    \subsection{Chemical Encoding is a Barrier to Developing Continuous Models}
    The standard method of finding an all-encompassing relationship in a dataset with hundreds of variables is through fitting the data on a neural network. Our work reveals that a neural network cannot be used to predict the relationship between a mass spectrograph and its corresponding chemicals due to its intractability: specifically, similar chemicals (chemicals that are structurally similar) have dissimilar spectra; therefore, there is no possible continuous loss function that a neural network can attempt to minimize to begin with. An in-depth analysis reveals that in order for a neural network to be able to predict any existing relationship, it needs to understand the output, and this necessitates that the output space (the names of the chemicals) follows a numerical encoding. There are various forms of encoding which can be used, all of which are problematic.
    
    Firstly, the name of the molecule is encoded into a 16 character ID called the InChIKey through the SHA-256 hash function. One encoding strategy is to use a one-hot encoding, in which a Boolean column vector of 0s has a 1 only at one specific position. However, this yields a space complexity of $O(2^{2^{16n}})\in O(2^{2^n})$ which is too large for a neural network. Another encoding strategy is to use an integer encoding by assigning a natural number to each unique chemical. However, the integer encoding assumes that numbers close to each other correspond to chemicals that are similar, which is not true. Finally, the only other possible means of encoding the chemical is to directly encode the structure of the chemical through a graph neural network. This is not a meaningful solution since the output of the graph neural network layer would be fed in as an input to the artificial neural network, both of which have incompatible architectural layers.

    \section{Starting from Scratch: Keeping Track of and Destroying Assumptions}
    
    \subsection{Principal Component Analysis (PCA)}
    We first try to predict the spectral relationship with a simplistic model that makes known assumptions using the popular PCA method \citep{pearson1901liii}. We assume that all the spectra are formed from linear combinations of orthonormal basis vectors, where the physical interpretation of each basis vector is an unknown atom or a functional group. This assumption is reasonable since compounds can be viewed as a combination of various atoms or functional groups. Therefore, we use Principal Component Analysis (PCA) \citep{pearson1901liii}, which is a dimensionality-reduction algorithm that re-derives the basis vectors for high-dimensional data. Specifically, we pass the spectrum to the PCA algorithm, in the hopes of finding which bases vectors can fully express the chemical spectrum.
    
    \begin{center}
    \includegraphics[scale=0.6]{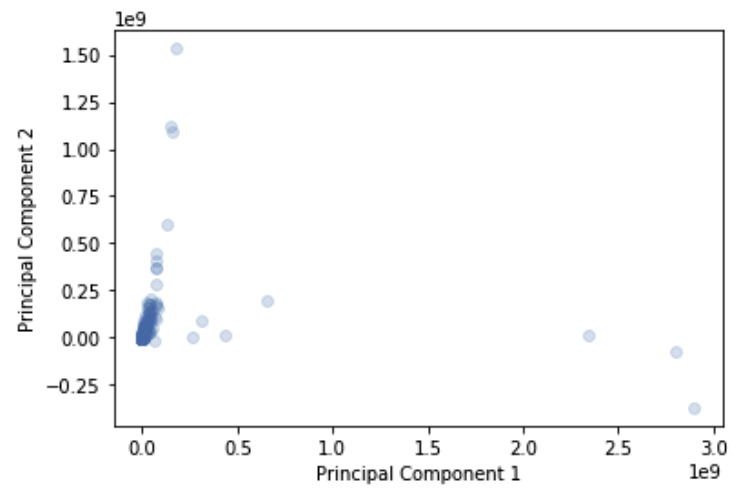}
    \includegraphics[scale=0.6]{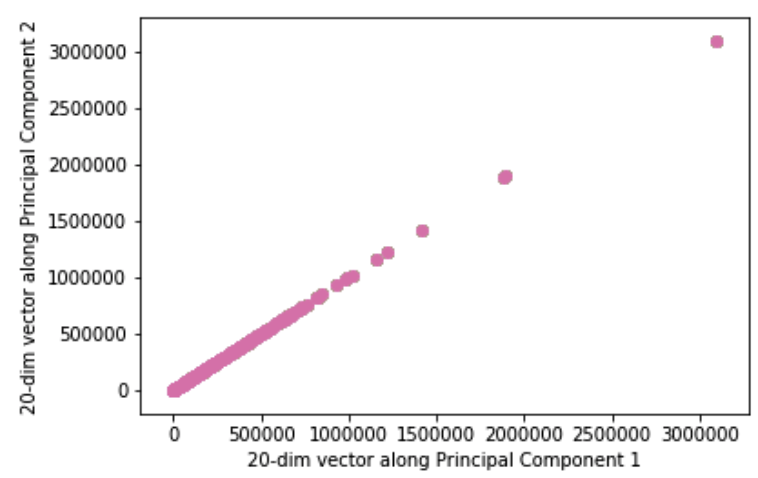}
    
        \textbf{Figure 3:} The first two dimensions of the 20-dimensional spectral matrix and its bases vectors generated from the PCA algorithm represented respectively. The points are generated from the Python Scikit-learn Decomposition-PCA library, and are scatter-plotted using the Python Matplotlib graphics engine.
    \end{center}
    
    However, two problems emerge from this approach:
    \begin{enumerate}
    \item The basis vectors for any dimension greater than 1 is not unique. A specific example of this is in $\mathbb{R}^{2}$ where the basis vectors for any set of vector is just any pair of orthogonal vectors. Therefore, when we try to reduce the 2660-dimension binned data to any reasonable size (smaller than 100), we extract orthonormal basis vectors, where none of the basis vectors correspond to any spectra. If we knew the rotation matrix that would superimpose the extracted basis vectors into a physically realizable solution, and we don't, this would still not solve the problem since the PCA-produced coefficients for each basis vector are very similar; therefore, the reconstructed compounds would be too distinct from the original compounds to be of any meaning. Hence, PCA does not provide a sufficient decomposition for the data.
    \item As a consequence of (1), the coefficients of some of the basis vectors produced by the algorithm can be negative. Since basis-vectors correspond to the presence of a specific substance, negative coefficients correspond to negative intensities which is not a physically valid solution.
\end{enumerate}

    \subsection{Non-negative Matrix Factorization (NMFA)}
    
    Since PCA fails to decompose the data, we move one assumption down the ladder and stop assuming that the spectra are formed from linear combinations of any orthonormal basis vectors, and are instead from linear-combinations of sparse orthonormal basis vectors. Furthermore, to make the solution physically meaningful, we assume that none of the coefficients are negative, since all peaks have to be positive.
    
    Therefore, we develop a non-negative matrix factorization (NMFA) model \citep{lee2000algorithms} which approximately factorizes the spectral matrix $X$ into a matrix of N columns of weights $W$ and N rows of basis vectors $H$. We apply the NMFA model to the spectrum to identify the weights and the basis vectors, while experimentally optimizing N in order to reduce N while maintaining a sufficient degree of intensity of the data. To optimize N, we need to decide if the reconstructed spectra is close enough to the original spectra. In general, verifying that the learned factorization reconstructs the original spectrum takes $O(N^3)$ time using naïve matrix products, up to bit complexity issues \cite{anand2024bit}. This can be improved using fast matrix multiplication in time $O(N^\omega)$ where $\omega<2.371$ \citep{alman2024refined}. 
    \begin{center}
    \includegraphics[scale=0.63]{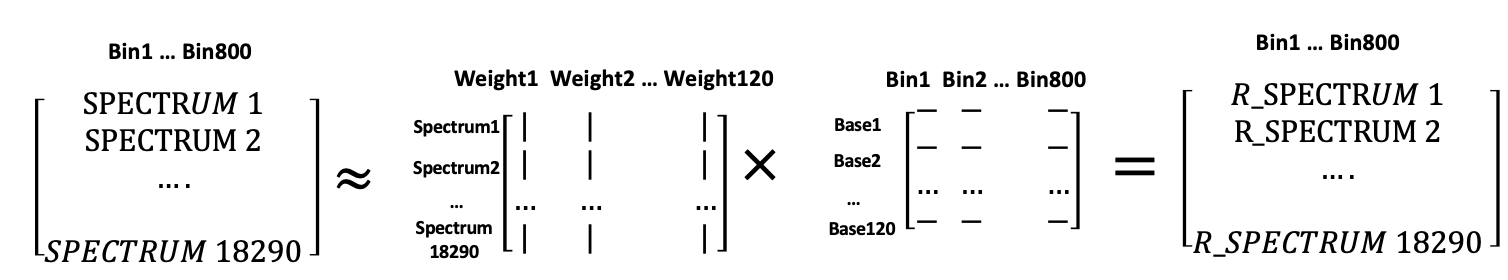}
    \textbf{Figure 4: Pipeline of Reconstructing Spectra}
    \end{center}
    
  To obtain a reconstruction of the original matrix using fewer principal components, we use the method described above. Note here that the product of the $i$'th column of W and the $i$'th row of H forms a reconstruction of the $i$'th spectrum. Using this process, we plot the sum of the absolute values of the differences of the spectral matrix between the original spectra and the reconstructed spectra (Figure 5).
    \begin{center}  
    \includegraphics[scale=0.69]{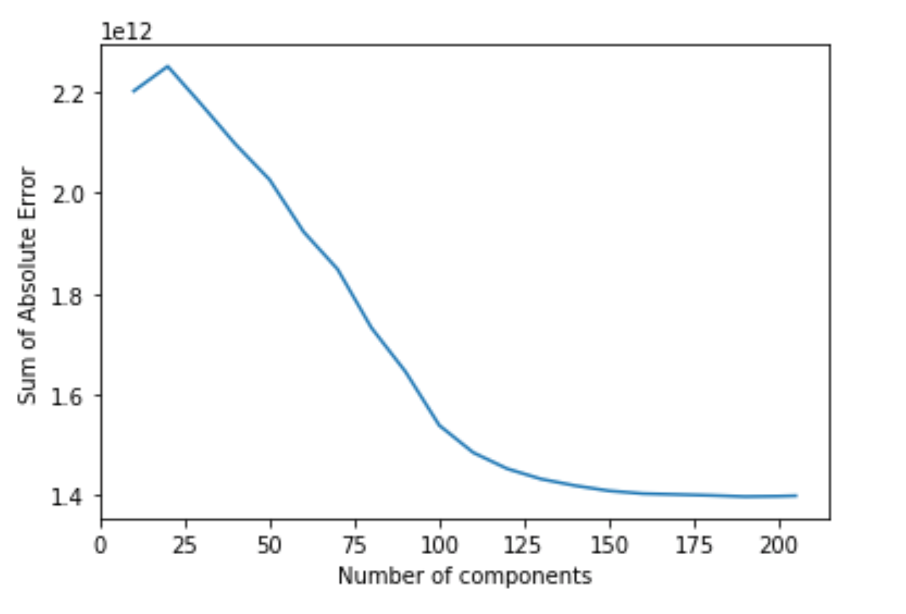}
    
        \textbf{Figure 5:} The sum of the absolute values of the differences of the spectral matrix between the original spectra and the reconstructed spectra, plotted against varying number of components.
    \end{center}

    Figure 5 demonstrates that at ~125 components, more components does not cause the NMFA error to decrease significantly. Therefore, we chose 120 components and ran the non-negative matrix factorization algorithm to generate the $H$ and $W$ matrices. We then plotted the actual spectrum and reconstructed spectrum for various substances and observed a variance in the level of accuracy of reconstructions. Figure 6 displays a relatively high-quality reconstruction of a sample. Throughout the reconstructions, we observed that reconstructed spectra which were not consistent with the actual spectra differed slightly in horizontal position, but vastly in vertical scaling.

    \begin{center}
    \includegraphics[scale=0.4]{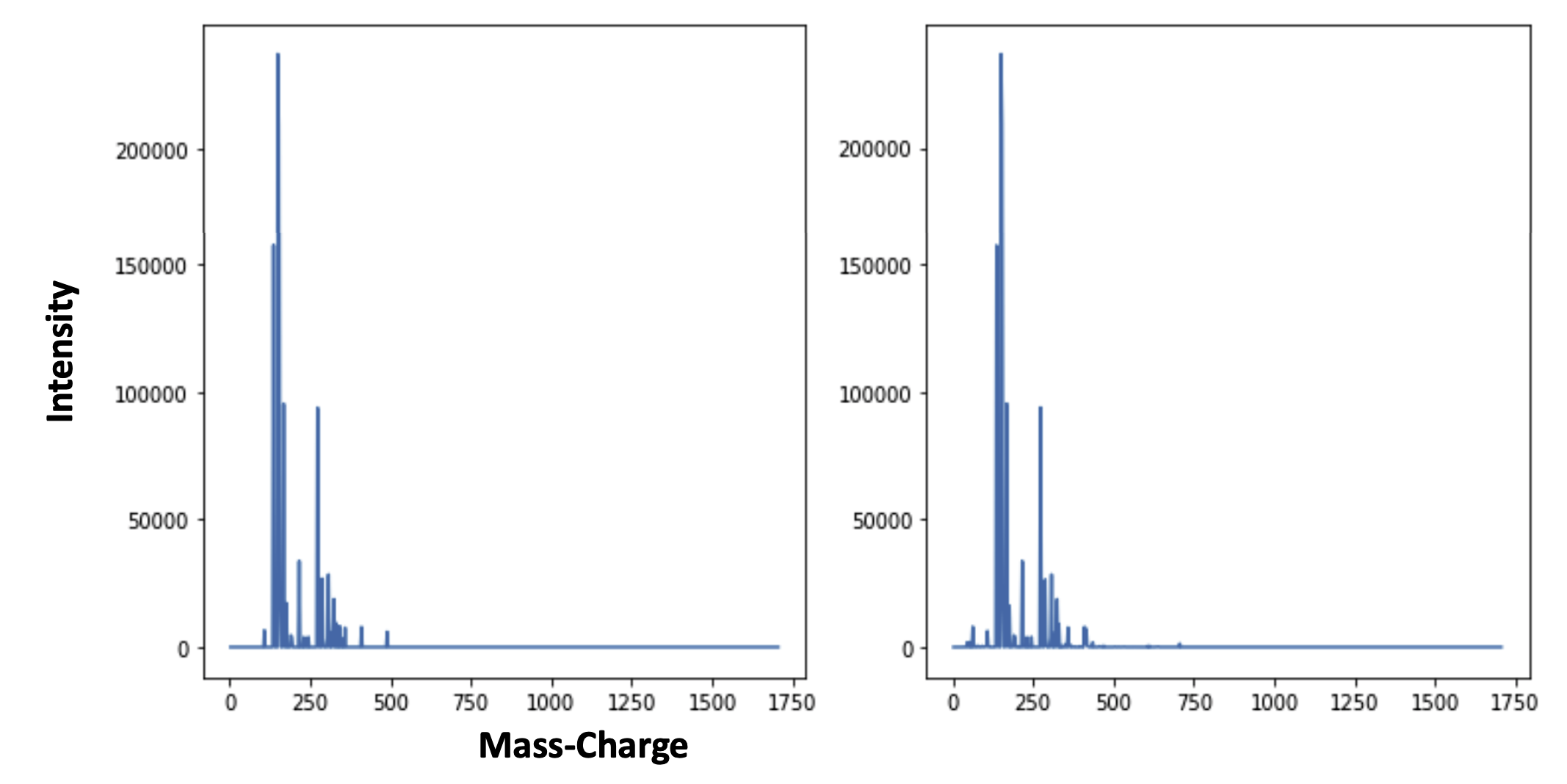}
        
        \textbf{Figure 6:} The actual spectrum (left) and reconstructed spectrum (right) of $C_{10}H_{16}N_{5}O_{13}P_{3}$. The reconstructed spectra was produced using Python's Scikit-learn NMFA library and plotted using Python's Matplotlib graphics engine.
    \end{center}

    We then visualize the actual and reconstructed spectral matrices produced by the non-negative matrix factorization (in Figure 7) and note that the differences in the reconstruction are (at a high-level point of view) negligible, as desired.
    \begin{center}
    \includegraphics[scale=0.85]{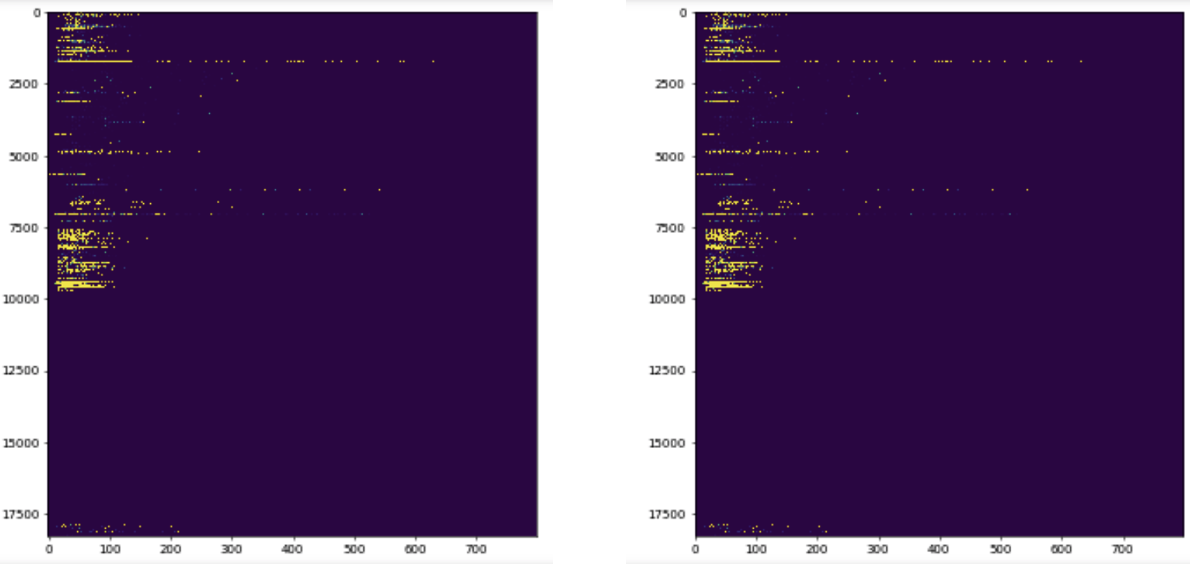}
    
    \textbf{Figure 7:} The actual spectral matrix (left) and reconstructed spectral matrix (right). The visualizations were produced using Python's Matplotlib graphic engine and the Python Imshow feature.
    
    \end{center}

 Additionally, we plotted the actual and reconstructed spectral matrix on a logarithmic plot (scaling each plot from the lowest values to the highest values of the corresponding spectral matrix). In doing so, the differences in the reconstruction became more distinct. \newline 
    
    \begin{center}
    \includegraphics[scale=0.7]{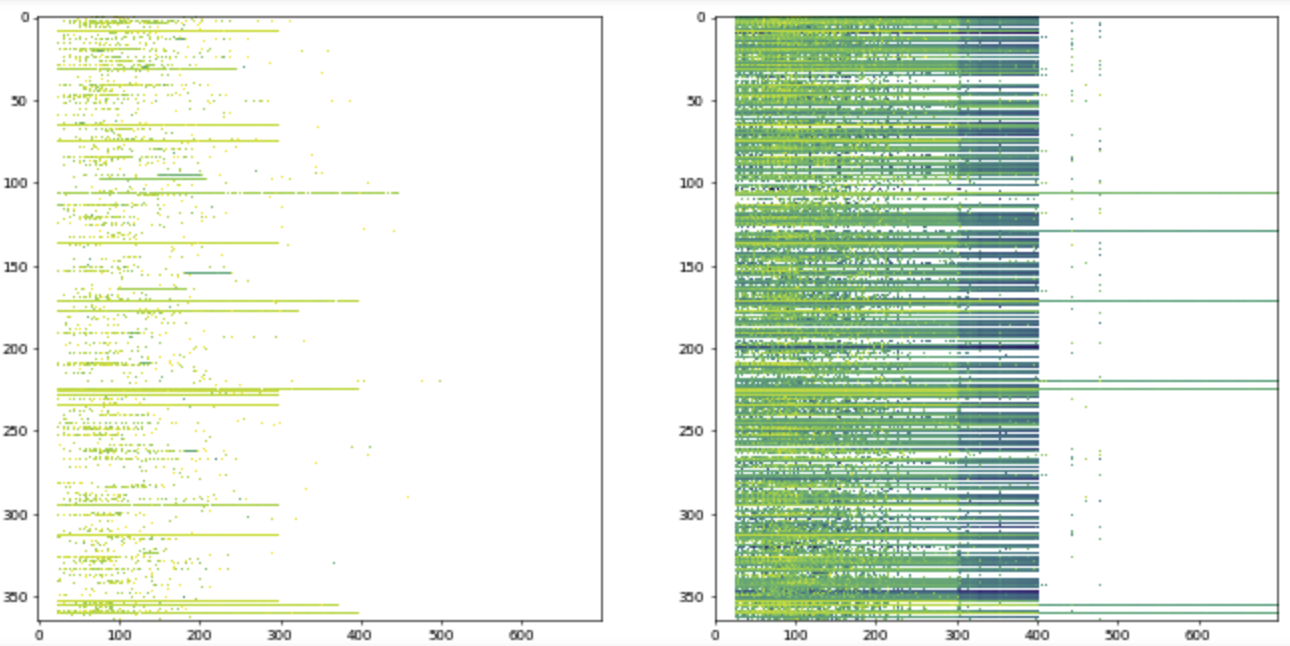}
    
    \textbf{Figure 8:} The actual spectral matrix (left) and reconstructed spectral matrix (right) plotted on a logarithmic scale.
    \end{center}

    \section{Extracting NMF Bases and Developing an Identifier algorithm}
    
    The ideal physical interpretation of each basis vector is that it represents some 'token', where each token is a monoatomic element or a functional group that is a building block of all chemical compounds. Therefore, if the NMFA decomposition correctly produced a set of tokens, they would be represented by the bases vectors. In this instance, the spectrum of the base would correspond to one significant peak that represented this smaller token. Since this is falsifiable, we randomly plotted 2 of our basis vectors (in Figure 9) and affirmed that the basis vectors indeed contained a significant peak which confirmed a correspondence to a small building-block-like structure. We additionally confirmed that this pattern propagated throughout all the basis vectors.

    \begin{center}
    \includegraphics[scale=0.64]{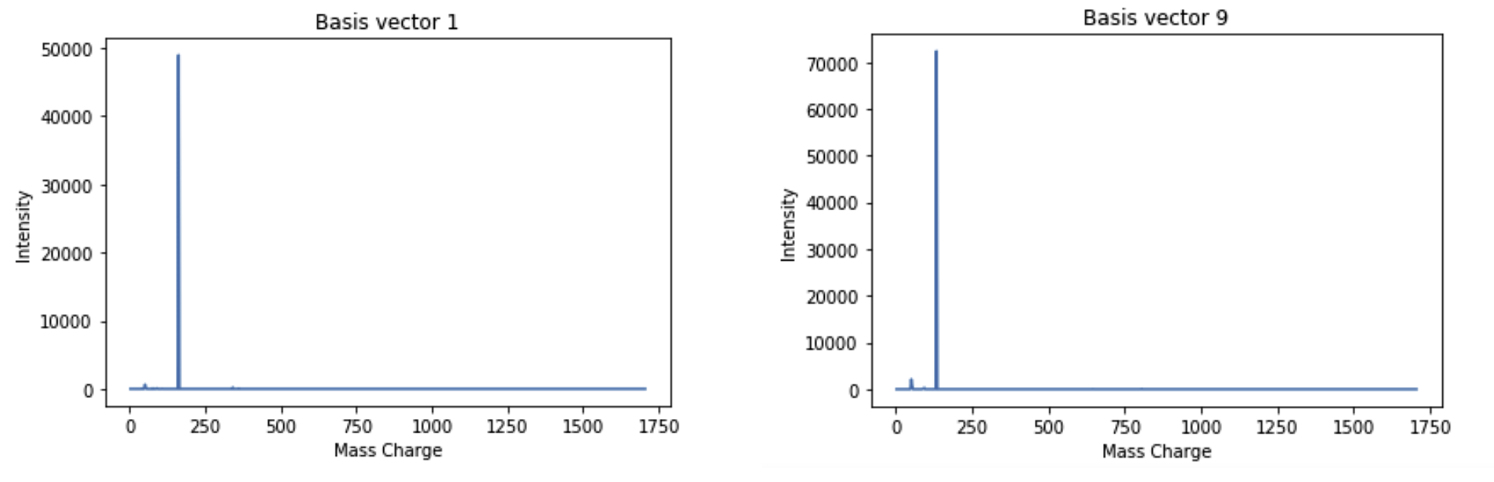}

    \textbf{Figure 9: }The first (left) and ninth (right) basis vector extracted from the weight matrix produced from the NMFA decomposition with 120 components.
    \end{center}

    \section{Extraction Algorithm}

Previously we showed that the NMFA extracted basis vectors that were consistent with possible tokens, where we defined tokens to be either monoatomic elements or small functional groups that could be possible building blocks of larger chemical compounds. However, we cannot naïvely know whether this is true without an in-depth analysis. \newline

Therefore, we developed the following algorithm (SEARCHER) that searches for various tokens in compounds to determine their mass-charge ratio. Since most compounds in the dataset were organic and were only ionized once, the mass-charge ratio is effectively the same as the atomic mass. Therefore, we can determine if the basis vectors are physically meaningful by comparing the atomic mass that the SEARCHER algorithm outputs with the known atomic mass of the token. In turn, this provides a better statistical test for checking that the basis vectors actually represent real physically interpretive tokens.

\begin{algorithm}
\caption{SEARCHER}
\begin{algorithmic}
\Require Token
\State \quad Tokenize all the compounds in the dataset and look for all the compounds containing the token.
\State \quad Find the reconstructed spectrum of each compound that contains the token.
\State \quad Average all the intensities in each bin across every compound that contains the token.
\State \quad Find the bin (mass-charge) that contains the highest-intensity.
\State \quad Return the mass-charge of the corresponding bin.
\end{algorithmic}
\end{algorithm}

\section{Results of the SEARCHER algorithm}

\subsection{Running SEARCHER on Oxygen}
When the searcher algorithm was run on the 'O' token corresponding to Oxygen, we obtained surprising results (Figure 10). Firstly, the algorithm converged on one specific basis-vector. When this basis-vector was examined, it contained a single peak that was primarily at the mass-charge ratio of 31.14505 which is exactly the atomic mass of Oxygen. Note that this result was obtained without ever telling the algorithm the atomic mass of any substance at any point in time, therefore conclusively showing that the basis vector has physical meaning in representing a token.

    \begin{center}
    \includegraphics[scale=0.7]{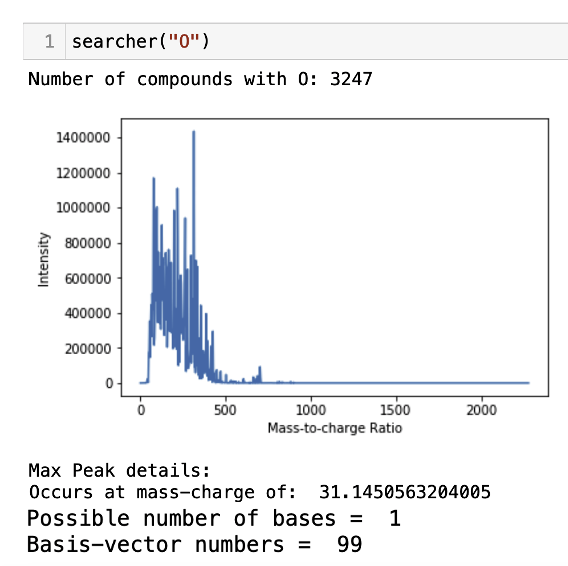}
    
    \textbf{Figure 10: }The output of the searcher algorithm on a 'O' token corresponding to Oxygen. The highest peak here occurs at the mass-charge of 31.145 which is to 2 s.f. the atomic mass of Oxygen.
    
    \end{center}

\subsection{Running SEARCHER on Chlorine}

We similarly ran the algorithm on the Chlorine token and obtained an atomic mass of 22.87 (Figure 11), whereas the atomic mass of the average Chlorine ion is 35.5 amu. To explain this, we notice that 22.87 is (to 2 s.f.) accurate to the atomic mass of Sodium, and that many of the data collectors in this dataset dissolved their compounds in a solution of brine (concentrated sodium chloride solution), where the sodium ions were likely picked up in the mass spectrometer. Thus, we believe the algorithm is still working correctly and that this is a fair result.

    \begin{center}
    \includegraphics[scale=0.5]{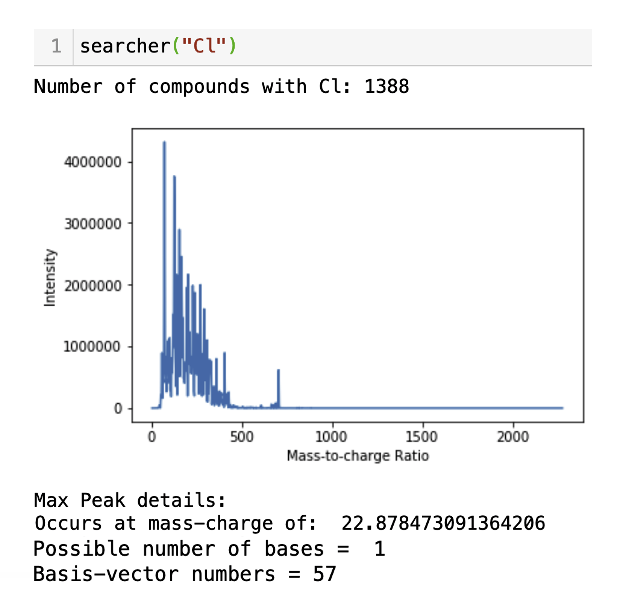}
    
    \textbf{Figure 11: }The output of the searcher algorithm on a 'Cl' token corresponding to Chlorine. The highest peak here occurs at 22.87 which is to 2 s.f. the atomic mass of Sodium. We suspect this is because all solutions in the dataset are dissolved in brine (concentrated Sodium Chloride) which interferes with the readings.
    \end{center}

\subsection{Running SEARCHER on Carbon and Hydrogen}

When running the algorithm on the Carbon token (Figure 12 - left) and the Hydrogen token (Figure 12 - right), we obtained the exact same results. In particular, the algorithm converged the Carbon and Hydrogen token to one basis vector with a corresponding mass-charge value of 31.42 (to 4 s.f.). We know this to be an erroneous result, since Carbon's atomic mass is 12.01 (to 2 s.f.) and Hydrogen's atomic mass is 1.01 (to 2 s.f.), and this is actually the same atomic mass produced when the SEARCHER algorithm is run on the Oxygen token. \\

We again suspect that the algorithm is working correctly, and we attribute these results to the kinds of data we are working with: ~85\% of the compounds in the dataset are organic (of which, ~70\% contain Oxygen). Specifically, these Oxygen-containing compounds dominate the average compound, and the mass-spectrum of Oxygen dominates those of Carbon and Hydrogen, which are again reasonable choices for tokens.

    \begin{center}
    \includegraphics[scale=0.6]{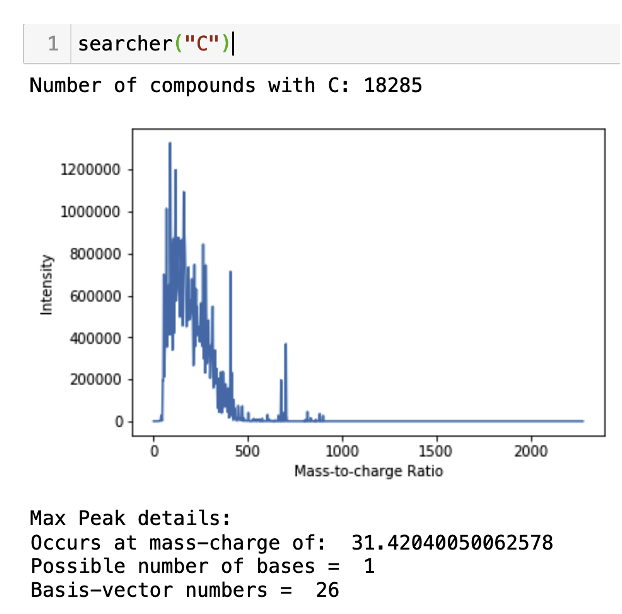}
    \includegraphics[scale=0.6]{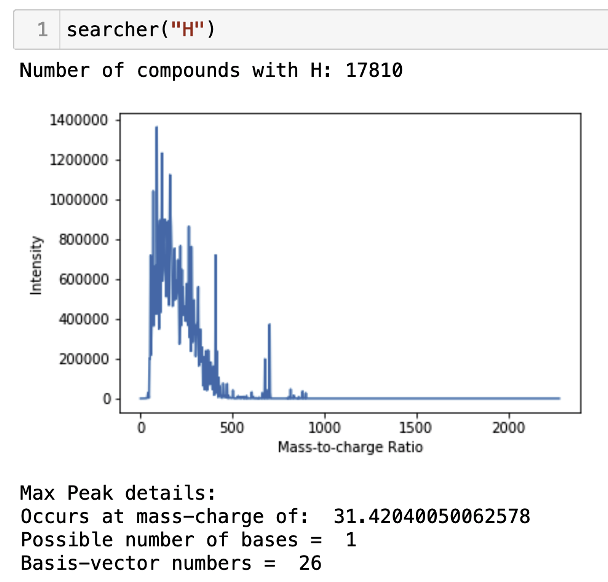}
    
    \textbf{Figure 12: }The output of the searcher algorithm on a 'C' token (left) corresponding to Carbon and a 'H' token (right) corresponding to Hydrogen. Note that these spectra are exactly the same, and that they are also congruent to the spectrum of Oxygen. This is because most compounds in the dataset are organic and contain Carbon Hydrogen and Oxygen atoms. However, the true Carbon and Hydrogen basis vectors are not revealed due to Oxygen's mass-charge ratio dominating those of Carbon and Hydrogen.
    \end{center}

It is non-trivial to rectify this problem (and other variations of it), since in the naïve approach, after every physical token is found and confirmed to be correct, we would have to adjust the spectral matrix by subtracting $c$(spectra of the token), $c\in\mathbb{R}$, to minimize the spectral peak corresponding to that token. However, finding $c$ is a hard problem since the spectra for most of these tokens contain smaller peaks which scale significantly with $c$, and since different spectra contain intensities that are of different orders of magnitudes, creating a statistical problem in determining the correct precision (the number of bits needed to encode the bytes) of $c$. \newline

Additionally, a clear computational bound which exists is that after each time the spectral matrix is adjusted and a token is determined, the NMF algorithm must be run again on the modified matrix, despite the extensive time costs of performing the NMFA. However, there might exist optimal methods to solve this problem that do not rely on aforementioned naïve approach and its uneconomic time complexity.

\section{Discussion}
\subsection{Potential Sources of Error and Noise}

The primary source of noise in this approach stems from the data. The European Massbank is a collection of numerous small datasets from various organizations. Each of these organizations had varying degrees of accuracy for their data collection, and each of them posted their data differently. For instance, Figure 13 displays the spectra for two different chemicals, each posted by different organizations. The second chemical (Figure 13 - top right) clearly conveys more information per peak than what is conveyed in the first chemical (Figure 13 - top left). However, by binning (discretizing the mass-charge regions of the data and letting each bin-value be the maximum intensity located within the region), we can eliminate the lack of standardization (as seen in the more comparable plots shown in Figure 13 - lower layer). While this does standardize the data sufficiently while allowing us to perform the essential task of comparing different spectra, binning does introduce a loss of information and there is no naïve way to overcome this problem.
\begin{center}
\includegraphics[scale=0.32]{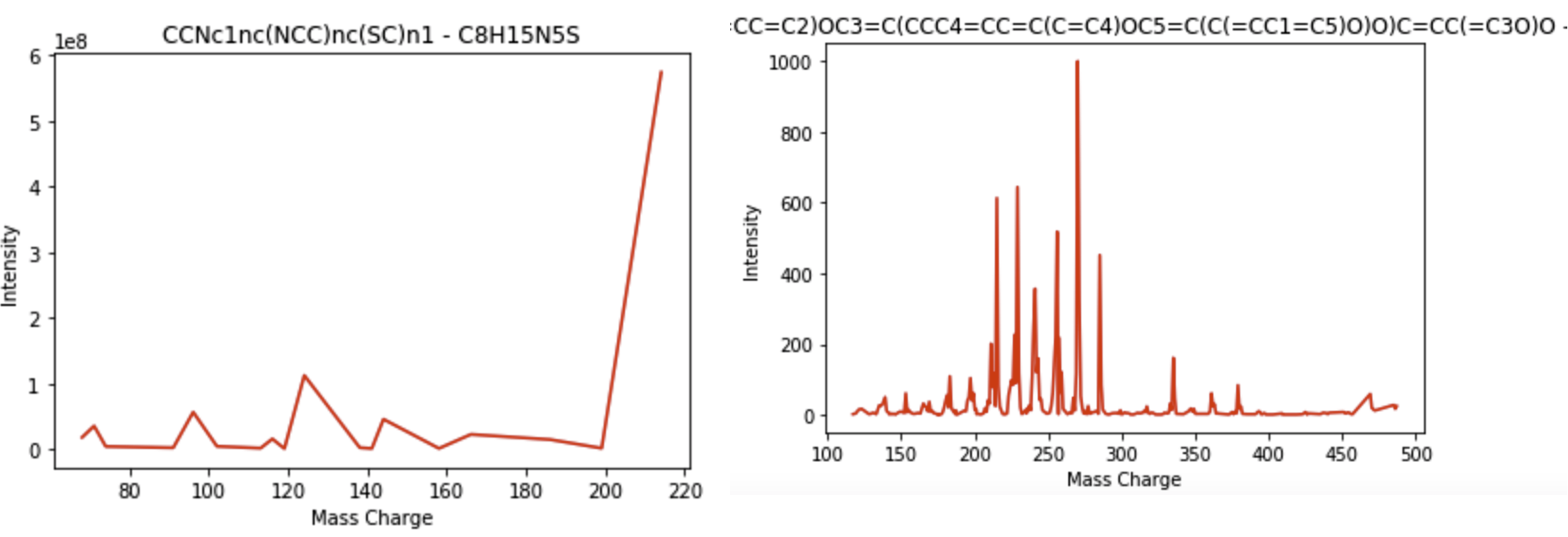}
\includegraphics[scale=0.5]{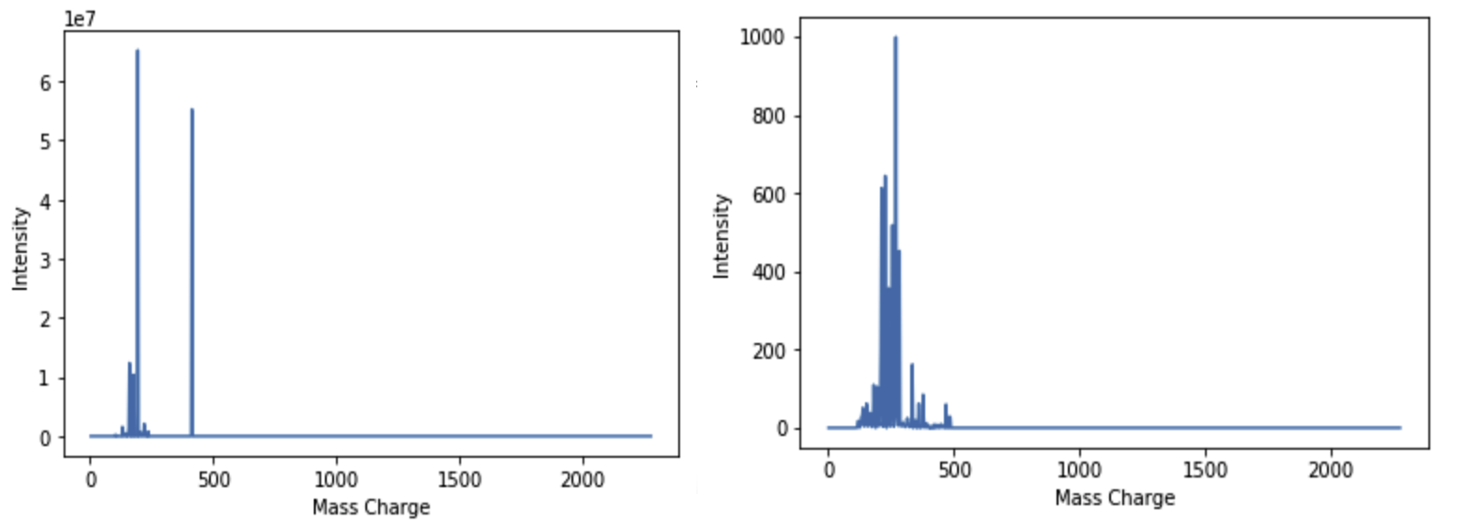}

\textbf{Figure 13:} Unstandardized data (top). Binned data (lower). Clearly, binning allows comparison between spectra. The mass-charge units are kg/C, and the intensity units are arbitrary (it is a measure of how hard different fragments collide with the detector and this measure is dependent on the detectors sensitivity) so only the relative amount is important.
\end{center}

One possible source of error in the data lies in the problem's chemistry. When a mass-spectrometer uses a sample, it vaporizes and bombards it with electrons to fragment the sample in various ways. These fragments represent the peaks in the mass-spectrograph. The mass-charge corresponds to the mass of the fragment by measuring how much the fragment deflects in the mass-spectrometer. It also measures the force with which the fragments collides with the detector, thus assigning the fragment an intensity value. The fragmentation within the mass-spectrometer is not necessarily unique since large compounds can fragment in many ways. Thus, the data collection groups repeat this process numerous times to obtain every possible fragmented source/peak, which introduces the possibility of missing fragments and repeated fragments at different mass-charge values due to random error/uncertainty\cite{DjoumbouFeunang2019,Xu2007}.

\subsection{Nonnegative Matrix Factorization Accuracies}

The non-negative matrix factorization (NMF) algorithm is mathematically equivalent to a clustering algorithm where the goal is to minimize $||X - WH||, H\ge0, W\ge0$. However, the algorithm (when used on our spectral matrix, $X$) could not minimize $||X - WH||$ to 0 when a suitable number of principal components (basis vectors) were used. Note here that 'suitable' is a twisted word. When the number of principal components is equal to 85582 (the number of compounds in the dataset, and therefore the number of rows in the spectral matrix), the minimization works perfectly since each spectrum gets assigned to a basis vector. However, in doing so, we learn nothing new and cannot extract any of the underlying information about the spectral matrix and its constituent building blocks. Therefore, the goal of the dimensionality reduction is to reduce the number of principal components as much as possible while still retaining as much accuracy as possible. Therefore, when we chose 125 principal components, we deliberately introduced some noise into the model. However, we also combat this by proposing that the token identification process of the 125 basis vectors be done manually to ensure that the results still make physical sense. This model has the potential to perform much better than any discrete model, and the results we have obtained are a testimony to this. The next section details the future work to be done on this project, as well as the problems that will need to be solved in order to identify the basis vectors. \newline

\section{Future Works}

To further solve this problem, future works could focus on understanding how to find residual bases vectors. In doing so, this would allow researchers to identify non-dominant basis vectors as physical tokens, thus avoiding the problem (and other analogs) of Oxgyen's spectrum dominating those of Carbon and Hydrogen. Furthermore, it is well-known that the presence of various atoms or functional groups has a large effect on the characteristics of the entire molecule. This effect is non-linear, since the presence of such a token changes the spectral landscape of the entire molecule. This cannot be accurately modeled by linear decomposition models such as PCA or NMFA, and might require novel decomposition methods, possibly through an appropriate modification of the Tucker decomposition process. Once the basis vectors are obtained, future work could focus on reconstructing all possible corresponding compounds from the known 'tokens' contained. Finally, an additional approach is to consider multi-agent RL approaches to model interactions between various molecules \citep{anand2024efficientreinforcementlearningglobal,anand2025meanfieldsamplingcooperativemultiagent}.

\section{Acknowledgements}
We gratefully acknowledge the following entities:
\begin{enumerate}
    \item Caltech SURF and the COSMIC Dawn Center - Niels Bohr Institute
    \item Professor Martin Hansen - Aarhus University, Denmark
    \item Xiaomin Zhu - Aarhus University, Denmark
    \item Vadim Rusakov - COSMIC Dawn Center, Niels Bohr Institute, University of Copenhagen, Denmark
    \item Professor Robert Phillips - California Institute of Technology, USA
    \item Human Metabolome Database and European Mass Bank Spectral Libraries
\end{enumerate}

\newpage

\bibliography{main}
\bibliographystyle{plainnat}

\end{document}